\documentstyle[11pt,shst2,epsf]{article}

\begin{document}

\title{Quasar Absorption Lines and the Intergalactic Medium}

\author{Buell T. Jannuzi}
\affil{National Optical Astronomy Observatories, P.O. Box 26732,
Tucson, Arizona 85726--6732}

\begin{abstract}
The importance of HST for the study of quasar absorption lines and of the
nature of the intergalactic medium is illustrated by reviewing
selected results from past HST observations. Topics reviewed include
the study of Ly$-\alpha$ absorbers at low redshift and the search for a
diffuse IGM at high redshifts.
\end{abstract}


\section{Opening the Ultra-violet Window}

Soon after quasars were recognized as extragalactic sources (Schmidt
1963; Greenstein and Matthews 1963; Schmidt 1965) it was pointed out
that as their light travels to Earth any intervening matter will leave
its imprint on the spectra.  This is true whether the intervening
medium is full of diffuse hydrogen causing a uniform decrease of the
quasar continuum (Gunn \& Peterson 1965; Scheuer 1965) or discrete
clouds producing separate absorption lines (by hydrogen, possibly
associated with galaxies, Bahcall \& Salpeter 1965; by hydrogen and
other species in galactic halos, Bahcall \& Salpeter 1966, Bahcall \&
Spitzer 1969).  However, even before the discovery of quasars, Lyman
Spitzer (1956) had pointed out the importance of UV spectroscopy for
understanding the physical conditions of the gaseous content of the
Galaxy, and by implication the halos of other galaxies and the gaseous
content of the universe in general.  As he pointed out, the majority
of the strong resonance absorption lines occur in the rest frame UV
(see Figure 1).
\begin{figure}[htb]
\centering
\plotfiddle{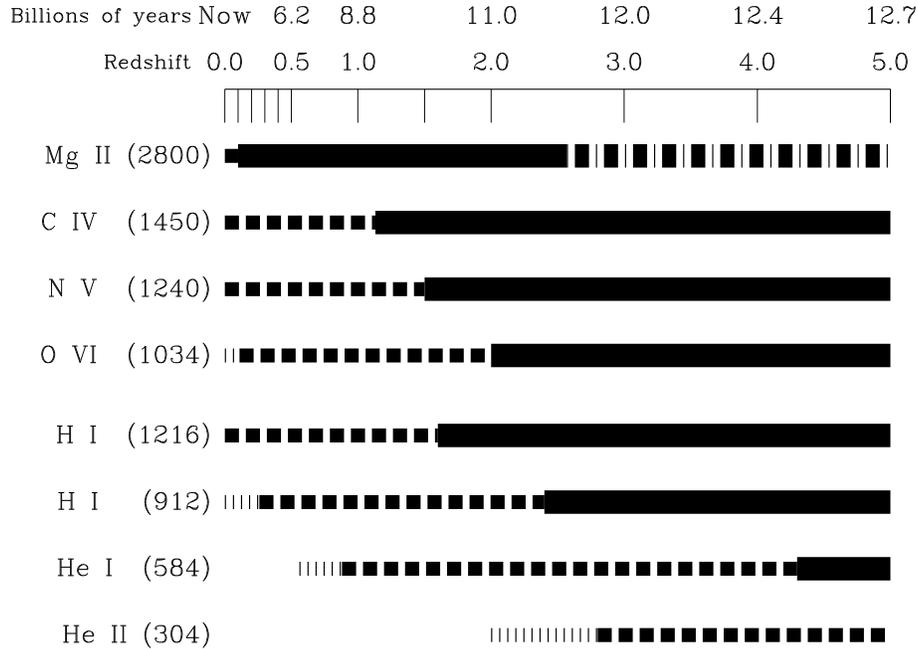}{3.1truein}{0}{80}{80}{-270}{-180}
\caption{The majority of the strong resonance absorption lines have
rest wavelengths which occur in the UV.  In this figure are plotted
the redshift ranges over which selected lines are observable in the IR
(applies only to Mg II), optical (solid line), ultra-violet
(observable with HST; dashed line), or far-UV (not observable with
HST, but observable with HUT, FUSE, or other UV missions; thin dashed
line).  Note that Galactic Lyman-limit absorption precludes any
instrument from being able to observe distant He~I or He~II absorption
below respective redshifts of 0.56 and 2.0.}
\end{figure}

At high redshifts intervening absorption systems can be studied from
the ground (for an example of the current state of the art see Figure
2). However, the gaseous content of the nearby universe and the far-UV
lines (e.g. He~II) occurring at high redshift can only be observed in
the UV.  Although HST is not the first telescope with UV sensitive
spectrographs, it is the first to provide both the spectral resolution
and the sensitivity to allow the extensive observation of the quasar
absorption lines.

Here is a list of three of the many important studies that the
observation of quasar absorption lines at UV wavelengths makes
possible:
\noindent 
1.) The evolution of the gaseous content of the universe can now be
traced by observing the changing number density per unit redshift of
Ly$-\alpha$ absorbers from the present (using HST, from $z_{\rm abs}
\approx 0.0$ to $z_{\rm abs} \approx 1.6$) back to when the universe
was 10\% of its current age (using groundbased telescopes like Keck to
observe the most distant quasars).  Such data are important for the
study of cosmology, star and galaxy formation, development of large
scale structures, and the composition of the ISM.  The dramatic 
changes that occur from high to low redshifts are illustrated by
comparing Figures 2-4.  Quantifying and understanding these changes in detail
is the continued focus of the HST quasar absorption line key
project and  other research efforts as well.

%
\begin{figure}[htb]
\centering
\plotfiddle{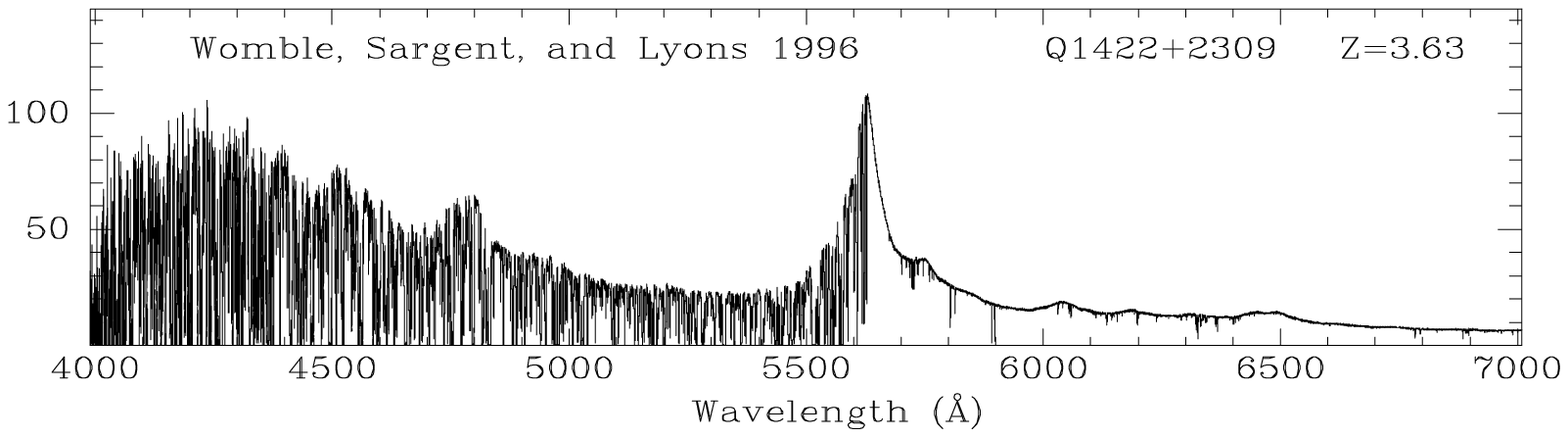}{1.25truein}{0}{93}{93}{-290}{-390}
\caption{The large number of absorbers blueward of Ly$-\alpha$, ``the
forest'', is quite evident in this Keck~I and HIRES observation of
the $z=3.63$ quasar Q1222$+$2309 obtained by Womble et~al. (1996).}
\end{figure}
\begin{figure}[htb]
\centering
\plotfiddle{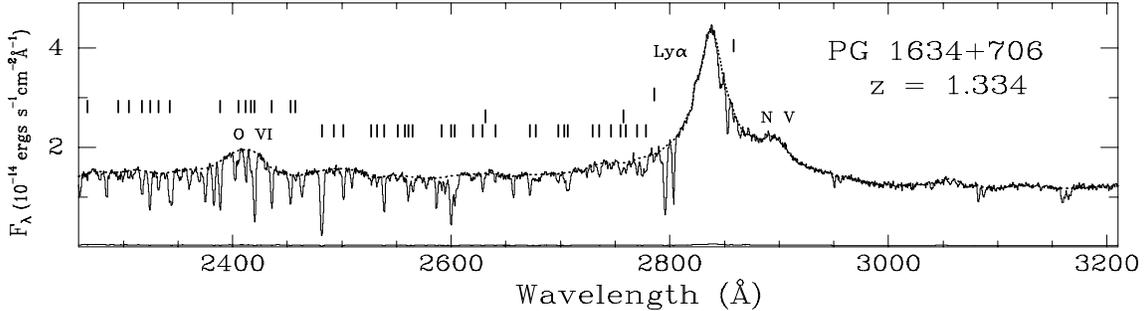}{1.27truein}{0}{93}{93}{-290}{-390}
\caption{The number of absorbers blueward of Ly$-\alpha$ is noticeably
less in this HST and FOS spectrum of PG~1634$+$706 (Bahcall
et~al. 1996).  Each Ly$-\alpha$ line is indicated with a vertical
tick mark above the location of the line.}
\end{figure}
\begin{figure}[hbt]
\centering
\plotfiddle{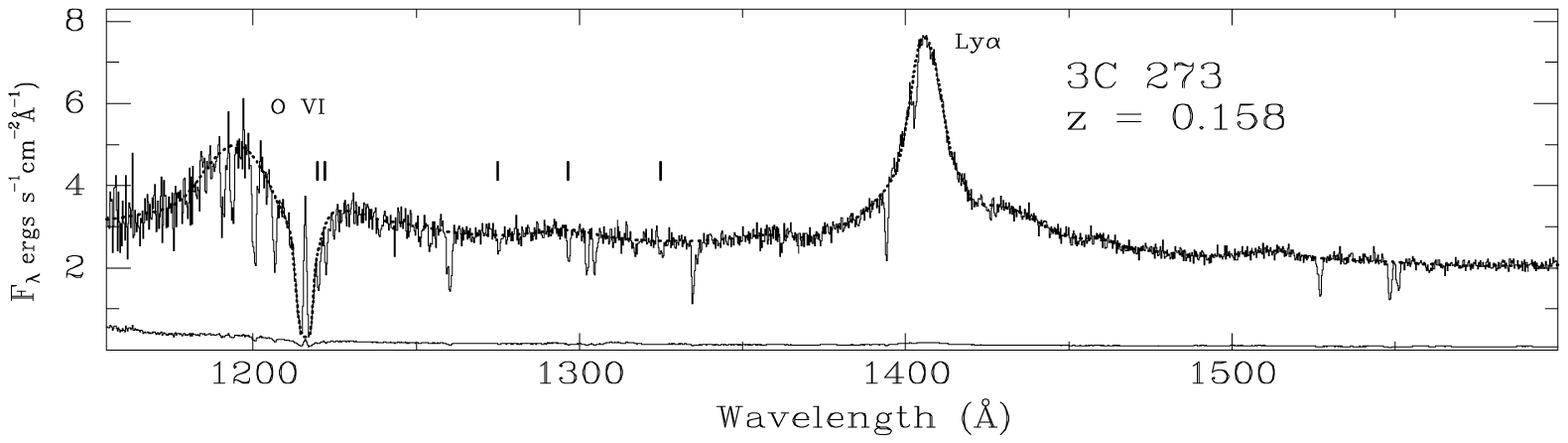}{1.25truein}{0}{93}{93}{-290}{-390}
\caption{By the low redshifts probed by this HST observation of the quasar
3C~273 (Bahcall et~al. 1991) the Ly$-\alpha$ forest seems to have been
completely removed.  However, the number of absorbers is still in
excess of a simple extrapolation of the evolution observed at
redshifts greater than 1.6 (Bahcall et~al. 1991; Morris et~al. 1991)
Each Ly$-\alpha$ line is indicated with a vertical tick mark above the
location of the line. The other absorption lines are caused by the ISM
of our own Galaxy.  Note the two Ly$-\alpha$ lines near zero redshift
at the approximate velocity of the Virgo cluster.}
\end{figure}

\noindent
2.) For low redshift absorbers it is now possible to study their
relation to individual galaxies, groups, or clusters -- i.e. test the
1969 proposition of Bahcall and Spitzer, ``that most of the absorption
lines observed in quasi-stellar sources with multiple absorption
redshifts are caused by gas in extended halos of normal galaxies.''

\noindent
3.) Although efforts to detect a diffuse intergalactic medium (IGM)
through continuous absorption by neutral hydrogen have failed, we can
now extend the search by looking for He~II absorption that may be
easier to detect since the lower level of the ionizing background at
short wavelengths means that the He~II fraction is larger than the
fraction of H~I.

In the following sections I will give examples of how HST has
contributed to each of these problems.  The examples are drawn from
the work of the HST quasar absorption line key project and a few other
groups to illustrate this progress, with apologies to the many other
researchers that have also contributed to the study of these and other
problems through the use of quasar absorption lines observed with HST.

\section{The HST Quasar Absorption Line Key Project}

\subsection{Design and Goals of the Survey}

The HST quasar absorption line survey was an HST key project for
cycles 1-3, with carryover observations extending into cycle 4.  Led
by John Bahcall, the survey had the ambitious goal of obtaining a
large and homogeneous catalogue of absorbers suitable for the study of
the nature of gaseous systems and their evolution (Bahcall et~al.
1993).  While the well known telescope problems in effect prior to the
servicing mission reduced the original scope of the survey, the key
project still successfully observed 89 quasars with the higher
resolution (R$=1300$) gratings of the Faint Object Spectrograph. A
small subset of the quasars were observed from 1150--3300 \AA, but the
majority were observed only between 2200--3300 \AA~ or 1600--3300 \AA,
depending on the redshift of the quasar.  Targets were selected to be
bright and have low Galactic extinction ($b>20$ degrees).  The
distribution of the targets is shown in Figure 5. Redshifts of the
observed quasars range between 0.25 and 2.0.  Details of the data
calibration and analysis can be found in Schneider et~al.~1993,
Jannuzi \& Hartig 1994, and Bahcall et~al.~1996.  In all of our
analysis (from line measurements to line identifications) we have
tried to remove subjective decision making from the process and
replace it with well tested algorithms implemented through computer
software.  This allows us to run the same software on simulated data
in order to improve our understanding of the limitations of both our
data and our analysis techniques.

\begin{figure}[htb]
\plotfiddle{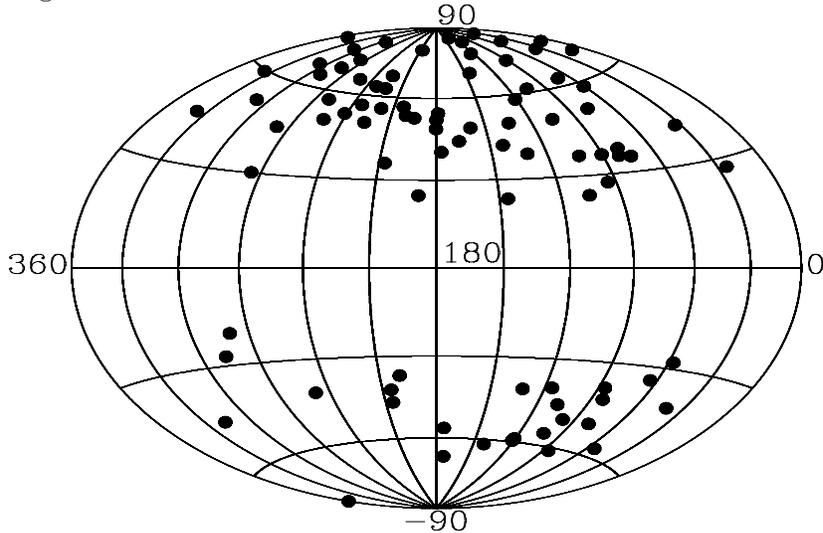}{2.35truein}{0}{70}{70}{-220}{-190}
\caption{The distribution in galactic coordinates of the 89 quasars
observed as part of the HST quasar absorption line survey.}
\end{figure}

\subsection{Past Results}

The first nice surprise that HST presented to us was a larger number
of low redshift Ly$-\alpha$ absorption systems in the spectrum of
3C~273 than might have been expected from a simple extrapolation of
the evolution in the number density (per unit redshift) of such
systems observed at high redshift (Bahcall et~al.~1991; Morris et
al.~1991).  Early results have also been produced by HST on the nature
and evolution of metal line systems (e.g. Reimers \& Vogel 1993;
Bergeron et~al.~1994).  Our understanding of the evolution of
Lyman-limit systems from high redshifts down to $z=0.4$ has been
improved (Storrie-Lombardi et~al.~1994; Stengler-Larrea et~al.~1995)
and a first attempt has been made to measure the proximity effect in
the spectra of low redshift quasars (Kulkarni \& Fall 1993).  

The key project catalogue of Ly$-\alpha$ absorbers makes it easier to
investigate the extent and nature of the relationship between
Ly$-\alpha$ absorbers and individual galaxies, groups, or clusters.
Many groups are actively working on this problem (e.g. incomplete or
single field surveys: Bahcall et~al.~1991, 1992; Morris et~al.~1993;
Spinrad et~al.~1993; to more extensive surveys in progress that have
presented partial results: Lanzetta et~al.~1995; Stocke et~al.~1995;
Le Brun et~al.~1995), but I have chosen to adapt a figure from Morris
et~al.'s (1993) study of the field of 3C~273 to illustrate both the
progress that has been made and how much more needs to be done
(Figure~6).  Despite a complete redshift survey of even the faintest
galaxies in the field, the study of the 3C~273 field gives a mixed
signal.  While some absorbers appear to be associated with the same
structures as the galaxies (as suggested by Lanzetta et~al., actually
part of the halos of the galaxies) other lines appear in voids (see
also Stocke et~al.) with no detected galaxy within 1 Mpc.  The Morris
et~al.~study is limited by the small number of Ly$-\alpha$ systems
along the line of sight toward 3C~273, resulting in a limited
comparison between the distribution of galaxies and absorbers.  In
fact no single line of sight provides enough Ly$-\alpha$ absorbers to
allow the accurate determination of the fraction of all absorbers
which are associated with galaxies or larger structures.  For some of
the other papers listed above the problem is similar or the galaxy
redshift surveys that they use are incomplete. Some of the other
surveys are also not able to address the relationship between the
absorbers and groups or clusters of galaxies because the galaxy survey
does not cover a large enough angular area to be able to identify a
cluster or group.  To determine accurately the fraction of Ly$-\alpha$
absorbers associated with galaxies and large scale structures requires
both the completion of the key project catalogue of absorbers and an
increase in the number of fields for which galaxy redshifts are
available (e.g. Sarajedini et~al.~1996).

\begin{figure}[htb]
\centering
\plotfiddle{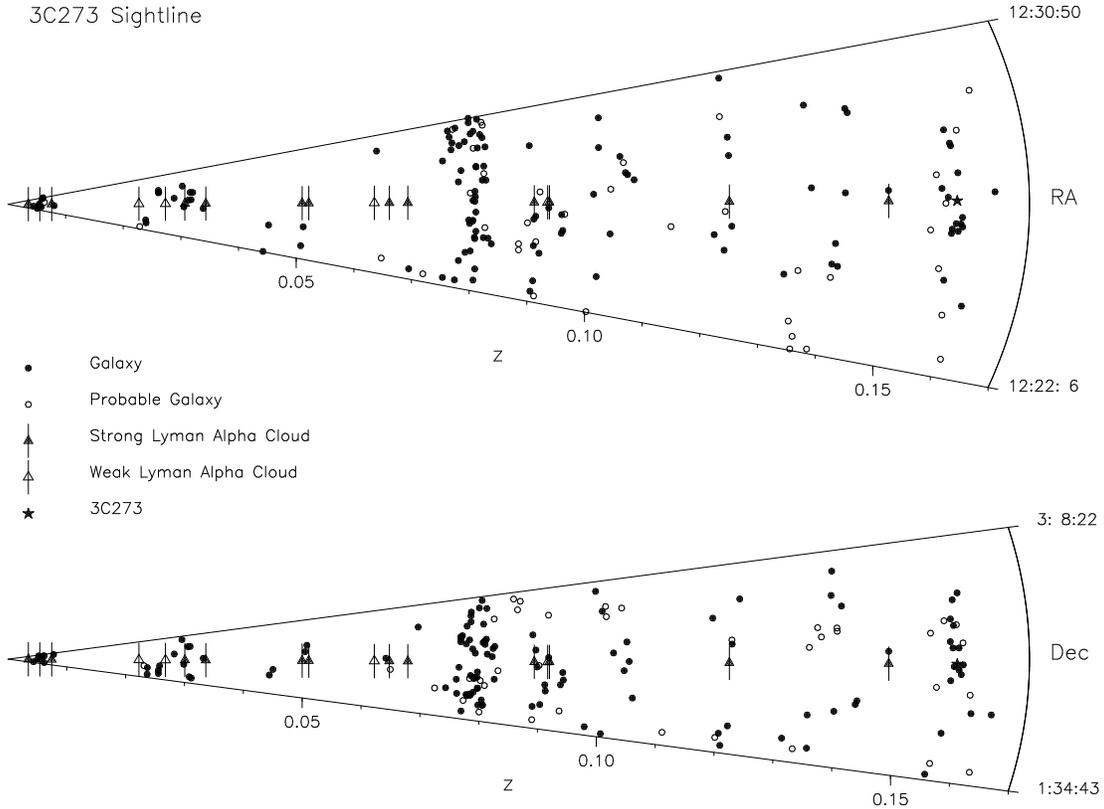}{4.0truein}{0}{88}{88}{-258}{-40}
\caption{Displayed in these pie-diagrams are the positions of the
galaxies in the field of the quasar 3C~273 and the locations of
Ly$-\alpha$ lines detected in the FOS and GHRS ultra-violet spectra of
3C~273.  Angles have been exaggerated by a factor of 15 to improve the
clarity of the figure, but results in a distorted plot with spherical
structures appearing elongated transverse to the line of sight. Note
that while some of the Ly$-\alpha$ absorbers appear associated with
galaxies, several have no detected galaxy within 1 Mpc.  Adapted from
Morris et~al.~1993, see their paper for complete discussion.}
\end{figure}

The same key project spectra have also provided valuable information
on our own Galaxy's halo and ISM (Savage et~al.~1993), the emission
line properties and spectral energy distributions of quasars (Espey
et~al.~1994; Weymann et~al.~1996; Laor et~al.~1994, 1995; Sulentic
this conference), and warm x-ray absorbers (in the quasar 3C~351,
Mathur et~al.~1994).

\subsection{Some New and Future Results}

A continuing focus of the key project is to determine the nature and
evolution of the low redshift Ly$-\alpha$ absorbers.  The number
density of such systems as a function of redshift is summarized in 
Figure~7 (see Bahcall et~al.~1996 for details).  At low redshift ($z<
1.3$), the key project data analyzed to date (about 10\% of the
expected final catalogue) is consistent with no evolution for $\gamma
= 0.58 \pm 0.50$ and $ dN/dz \propto (1+z)^{\gamma} $. This result is
derived from a maximum likelihood estimation for the observed lines in
those spectral regions where the 4.5 $\sigma $ detection limit is less
than 0.24~\AA.  We further find that the slope of the
observed low-redshift $dN/dz$ relation differs at the $2-4.5\sigma$
level of significance from the slope deduced from various ground-based
samples that refer to redshifts $z > 1.6$ (Lu et~al.~1991; Press et
al. 1993; Bechtold 1994).

\begin{figure}
\plotfiddle{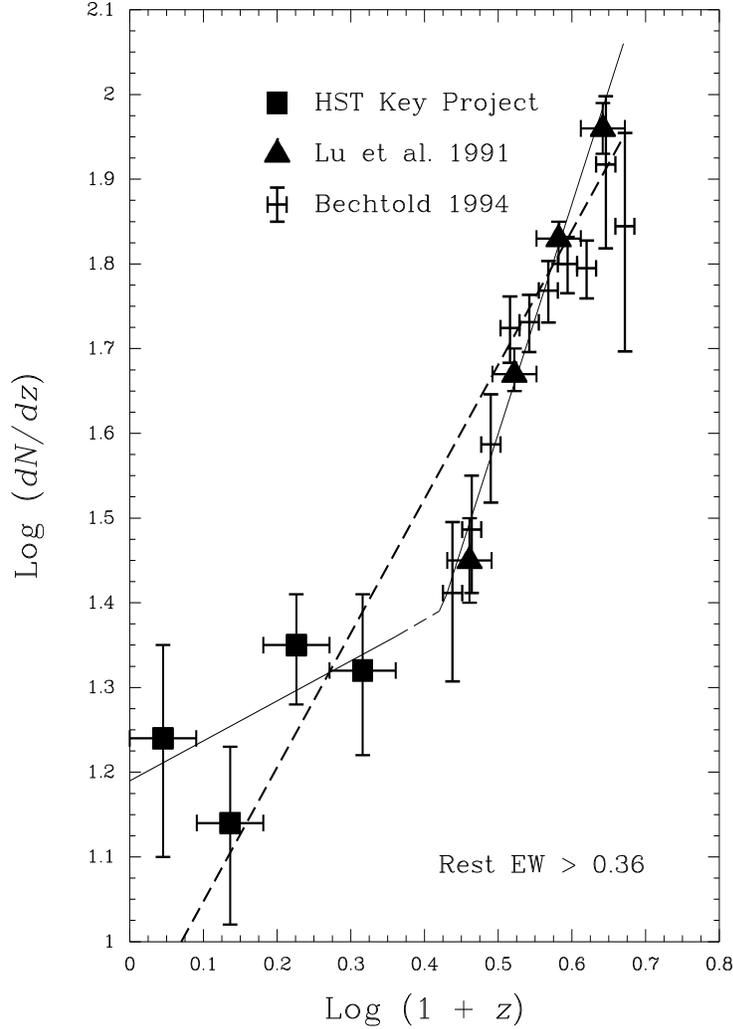}{4.9truein}{0}{80}{80}{-220}{-98}
\caption{The number density of Ly$-\alpha$ absorption systems is shown as a
function of redshift.  At redshifts below 1.6 the data come from
Bahcall et~al.~1996, including approximately 10\% of the final HST key
project catalogue of Ly$-\alpha$ absorbers.  At higher redshifts two
samples (Lu et~al.~1991; Bechtold 1994) are plotted which give
similar but not identical results. The separate fits to the HST and Lu
et~al.~data are shown as the two solid lines, with slopes of
$\gamma=0.48\pm0.54$ and $2.68\pm0.27$ respectively. The dashed line
shows the best fit to a single power law for both the HST and the Lu
et~al.~data and has a slope of $1.58\pm0.13$. A KS test indicates that
this fit is only acceptable at the 2.7\% level. Fits using the
Bechtold (1994) data give similar results although a single power law
fit is not ruled out as strongly. The fits were done using a
maximum likelihood technique on the unbinned data. The binned data
have been placed on the figure for reference purposes only.}
\end{figure}

As the number of absorbers in the analyzed catalogue increases it
becomes possible to study the clustering properties of Ly$-\alpha$
absorbers.  While we have yet to detect any signal in the two-point
correlation function, we have found evidence that about half of the
extensive metal line systems seen at redshifts between 0.4 and 1.3 are
accompanied by highly-clustered clumps of Ly$-\alpha$ lines which are
physically associated with the metal-line systems (details in Bahcall
et~al.~1996).

Our understanding of both the redshift evolution of all absorption
systems and of their clustering properties will improve as we complete
the catalogue of absorption systems.  The last observation of the key
project was made in May of 1995.  At the time of this meeting all of
the quasar spectra have been reduced, lines measured, and the lines
are being identified. The key project results I have reviewed have
been based on only part of the total absorption line data set (see
Figure 8). While we have analyzed one sixth of the objects, the
remaining five sixths include most of the higher redshift objects and
four fifths of the observed redshift path length.  Expected
improvements upon completion of the catalogue include: 1.)  examining
the evolution of Ly$-\alpha$ systems not only as a function of
redshift, but also as a function of neutral column density and 2.)
confirming or refuting the preliminary evidence for clustering of
Ly$-\alpha$ absorbers around metal line systems.

\begin{figure}
\plottwo{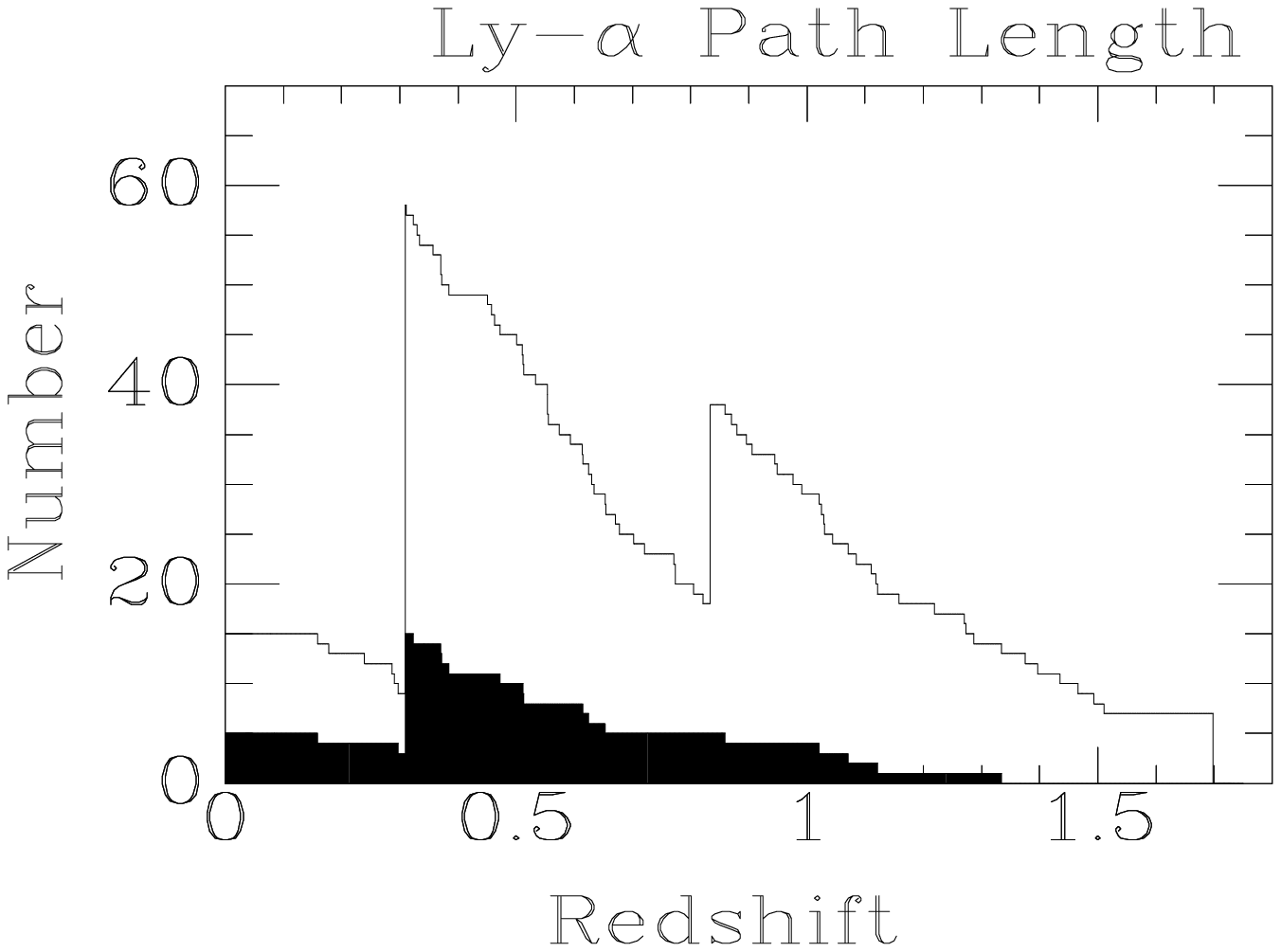}{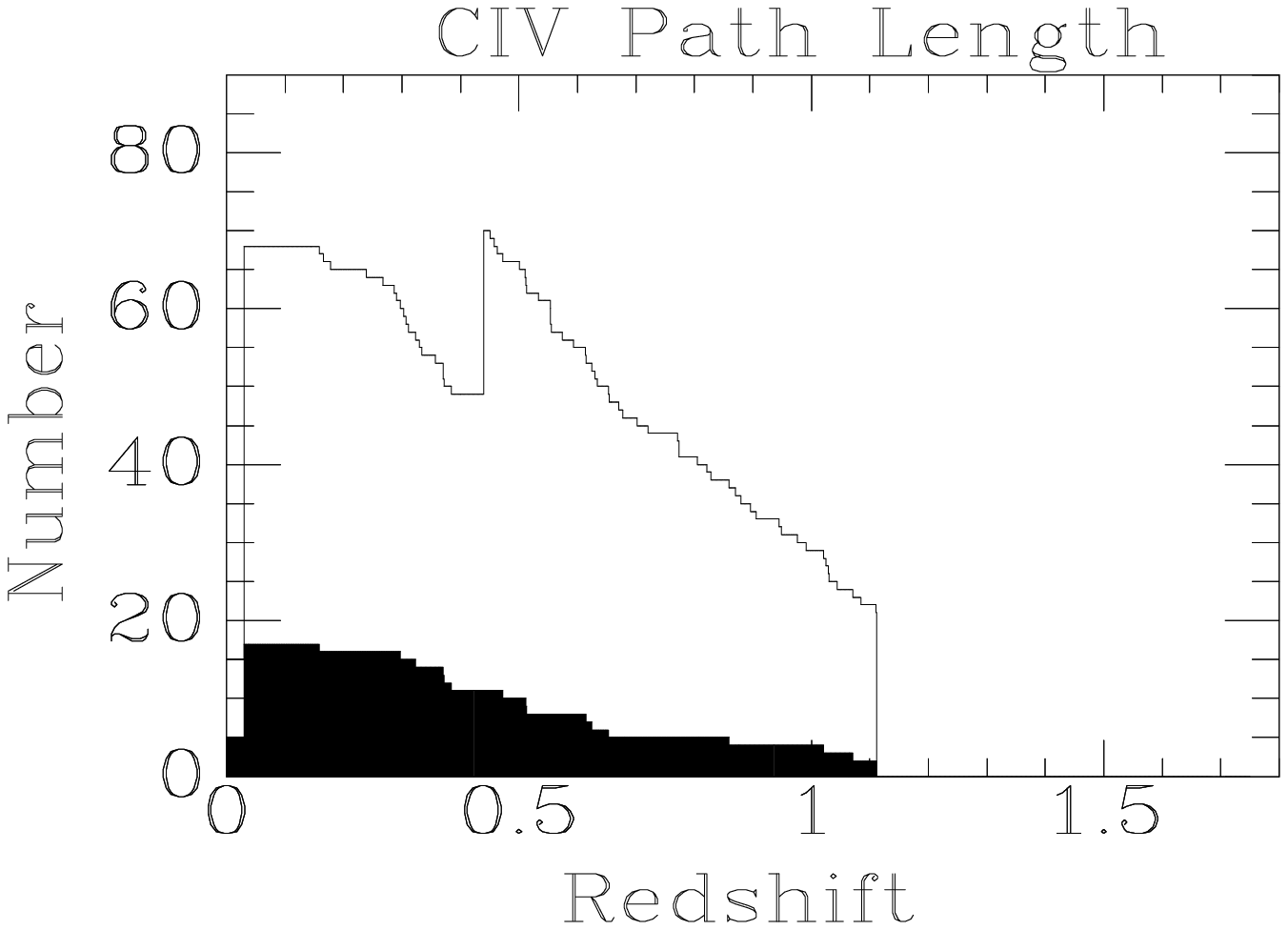}
\caption{Shown here are histograms depicting the number of quasars
observed as part of the absorption line key project which will
contribute information at each redshift.  The solid region indicates
the number of quasars whose data have been included in the published
papers and the outlined region indicates the complete sample currently
being analyzed. When the analysis is complete the size of the
absorption line catalogue will increase by a factor of 10, greatly
improving the determination of the evolution depicted in Figure 7.}
\end{figure}

\section{Has the IGM Finally Been Detected?}

We now leave the universe at low redshift behind and examine the
Herculean efforts that have been made to detect the diffuse
intergalactic medium with HST (a complete and detailed account of this
exciting area can be found in the contribution to these proceedings by
Dr. Jakobsen).  Excluding the detection of absorption assigned to weak
individual ``clouds'' (the low column density end of the Ly$-\alpha$
forest; the Bahcall-Salpeter effect), all efforts to detect absorption
by diffuse neutral hydrogen (the Gunn-Peterson effect) at ANY redshift
have failed.  The ionizing background radiation reduces the fraction
of H~I and He~I (Sargent et~al.~1980) and removes them as probes of
the diffuse IGM (note, He~I has been observed in high column density
systems, the first detection being made with HST by Reimers \& Vogel
1993).  The lower ionizing background at short wavelengths might leave
a higher fraction of He~II and provide a means of detecting the
diffuse IGM, but the short wavelength of He~II (304 \AA) means that it
can only be observed at high redshifts.  This means that ``clear''
quasars, as Jakobsen call them, must be found.  Such quasars must be
bright, have redshifts greater than 3 (to have He~II observable with
HST), and be free of significant absorption from intervening
``clouds'', particularly the high column density systems whose
Lyman-Limit absorption would preclude the observation of He~II.
Jakobsen et al. (1994) and Tytler et~al.~(personal communication) both
conducted searches during cycles 1-3 using HST and respectively the
FOC and FOS to find candidate ``clear'' quasars.  In cycle 4 both
groups succeeded in detecting He~II absorption in the spectra of
distant quasars.

While the details can be found in Dr. Jakobsen's contribution, here
are some bottom lines.  There are now three detections of absorption
due to He~II. Two quasars (Q0302$-$003 and PKS1935$-$692) exhibit
black, continuous absorption blueward of the expected wavelength of He~II
at the redshift of the observed quasars ($z > 3$ for the two observed
with HST, see figures in Jakobsen et~al.~1994 and Jakobsen's
contribution; Jakobsen and Tytler 1996). The lower limits on the
optical depth of He~II absorption are 1.7 in both cases. A third
quasar, HS1700$+$6414, was successfully observed by HUT (Davidsen
1995) and shows He~II absorption beginning at $z=2.7$, but the
absorption is not as strong as at the higher redshifts observed with
HST.  It appears that both the HST and HUT observations can be
interpreted as consistent with each other given the possible evolution
between redshifts of 3.2 to 2.7 (Jakobsen, this meeting).

One problem with any search for a diffuse component in the IGM is that
as we are able to detect and resolve lower column density systems we
remove absorption from the previously unresolved ``diffuse'' component
and move it into the ``cloud'' component.  Q0302$-$003 has been
observed with Keck and the HIRES echelle spectrograph and a population
of very low column density Ly$-\alpha$ clouds has been detected by
Songaila et~al.~(1995) and they report that the detected population is
extensive enough that it is possible to explain the observed He~II
absorption without invoking a diffuse IGM. Neither the HUT nor the HST
FOC and FOS observations have the spectral resolution necessary to
distinguish directly between the He~II ``forest'' and a more diffuse and
uniform absorption.  The issue is likely to remain unsettled until the
existing lines of sight (or additional new detections, hopefully with
brighter background quasars) are successfully observed at a high
enough spectral resolution that the He~II associated with the hydrogen 
forest clouds can be resolved.

There is a second complication.  Cosmological simulations of the
universe at intermediate and high redshifts (e.g. Katz et~al.~1996)
indicate that we should now expect a complex distribution for the gas
in the IGM with filamentary structures covering a large range of
physical scales and conditions. There might not exist any component
that matches our expectations of a smooth or uniform ``diffuse IGM''.
It might be that the distinction between numerous, closely packed,
very diffuse (low column density) ``clouds'' and a more uniform
diffuse medium is purely a question of semantics, but the resolution
of this issue has implications for a variety of issues, including
understanding the physical conditions that exist during the formation
of galaxies (see Jakobsen, these proceedings for further discussion).

\section{End Matters}

The 1990's is the epoch of two revolutions in the the study of quasar
absorption lines.  Prior to HST and the Keck telescopes, quasar
absorption lines have been discussed and studied in distinct
subgroups, roughly separated by column density. At the extremes were
the Ly$-\alpha$ forest lines that were observed to be unclustered and
possibly composed of primordial material (based in part on the lack of
any detected metal line absorption) and the damped systems with their
high column densities and large gas masses identified as the
progenitors of spiral galaxies (e.g. Wolfe 1988).  Such divisions,
while still useful, are getting fuzzy as new results rapidly blur
distinctions.  Just one example (of many) is the detection of weak CIV
absorption associated with some fraction of low column density
Ly$-\alpha$ absorbers, systems that would have previously been
securely identified as part of the primordial ``forest clouds''
(e.g. Cowie et~al.~1995; Womble et~al.~1996). Such wonderful
observations require modification of the pre-HST-Keck picture of
absorption line systems.  How should we modify the old ``standard
picture''? I am not sure.  But I do think that a second revolution is
going to provide critical guidance in the development of the new more
complex and detailed models.  The second revolution is the progress
theorists have made in leaving behind spherical cloud, slab, or
mini-halo models and replacing them with the help of super computers
to generate full hydrodynamic and SPH simulations of the evolution of
the universe. Three groups are now able to not only generate
simulations of large scale structures, but also simulated quasar
absorption line spectra along numerous lines of sight through their
simulations that can be compared to real observations (see Zhang,
Y. et~al.~1995; Hernquist et~al.~1996 and Katz et~al.~1996; Cen
et~al. 1994, Miralda-Escud\'e et~al.~1996). The challenge ahead is to
extract the best set of observables from both the simulations and the
various data sets so that cosmological models might be discriminated
against. Furthermore, enough simulations (and observational data!)
need to be generated that the uniqueness of ``good fit'' models can be
tested.

In his introduction to the Hubble Deep Field project, Bob Williams
ably described how HST has opened up the distant universe to our view.
He speculated that one of the Hubble Space Telescope's lasting and
important legacies would be providing us our first ``clear'' images of
the early history of the universe.   In the future HST will also be
remembered for making possible unique studies of the more evolved and
nearby universe.  WFPC-2 is providing exquisite images of galactic
sources and nearby objects that reveal a wealth of previously
unobservable detail (see for examples the contributions to these
proceedings by Bally, Livio, Machetto, and O'Dell). But HST should
also be remembered for the unique information provided by its
spectrographs.  By making it possible to study quasar absorption lines
in the ultra-violet HST has already provided important data
about the gaseous content of the universe at both low and high
redshifts. This legacy will continue to grow as existing data is
further analyzed and when STIS makes its appearance on HST.

\acknowledgments
I thank Jill Bechtold, Simon Morris, Donna Womble and Wal Sargent, and
the HST quasar absorption line key project team for providing data
used in the figures and acknowledge valuable discussions with Jill
Bechtold, Peter Jakobsen, David Weinberg, and the entire HST quasar
absorption line survey key project team. Hans-Walter Rix and David
Weinberg provided useful comments on an early version of this paper. I
thank the meeting organizers, particularly the local organizers, for
managing to host such an enjoyable conference given the logistical
problems brought about by the Paris transit strike.


\begin{references}


\reference Bahcall, J. N., Jannuzi, B. T., Schneider, D. P., Hartig,
G. F., Bohlin, R., \& Junkkarinen, B. 1991, \apj, 377, L5
\reference Bahcall, J. N., Jannuzi, B. T., Schneider, D. P., Hartig,
G. F., \& Green, R. F. 1992, \apj, 397, 68
\reference Bahcall, J. N., et~al.~1993, \apjs, 87, 1
\reference Bahcall, J. N., et~al.~1996, \apj, 451, 19
\reference Bahcall, J. N., \& Salpeter, E. E. 1965, \apj, 142, 1677
\reference Bahcall, J. N., \& Salpeter, E. E. 1966, \apj, 144, 847
\reference Bahcall, J. N., \& Spitzer, L. 1969, \apj, 156, L63

\reference Bechtold, J. 1994, \apjs, 91, 1

\reference Bergeron, J., et~al.~1994, \apj, 436, 33

\reference Cen, R., Miralda-Escude\'e, J. Ostriker, J. P., \& Rauch,
M. 1994, \apj, 437, L9

\reference Davidsen, A. 1995, B.A.A.S., 186, 30.01

\reference Greenstein, J. \& Matthews, T. 1963, Nature, 197, 1041

\reference Hernquist, L. Katz, N., Weinberg, D. H., \&
Miralda-Escud\'e 1996, \apj, 457, L51

\reference Jannuzi, B. T., \& Hartig, G. F. 1994, in Calibrating
Hubble Space Telescope, ed. J. C. Blades \& S. J. Osmer (Baltimore:
STScI), 215
\reference Jannuzi, B. T. et~al.~1996, in preparation

\reference Jakobsen, P., Boksenberg, A., Deharveng, J. M., Greenfield,
P., Jedrzejewski, R., \& Paresce, F. 1994, Nature, 370, 35
\reference Jakobsen, P. \& Tytler, D. 1996, in preparation

\reference Katz, N., Weinberg, D. H., Hernquist, L, \&
Miralda-Escud\'e, J. 1996, \apj, 457, L57
\reference Kulkarni, V. P., \& Fall, S. M. 1993, \apj, 413, L63

\reference Laor, A. et~al.~1994, \apj, 420, 110
\reference Laor, A. et~al.~1995, \apjs, 99, 1

\reference Lanzetta, K., Bowen, D. V., Tytler, D., \& Webb, J. K.
1995, \apj, 442, 538

\reference Lu, L., Wolfe, A. M., \& Turnshek, D. A. 1991, \apj, 434, 493

\reference Le Brun, V., Bergeron, J., \& Boisse\'e, P. 1995, A\&A, in press

\reference Mathur, S., Wilkes, B., Elvis, M., \& Fiore, I. 1994, \apj,
434, 493

\reference Morris, S. L., et~al.~1993, \apj, 419, 524
\reference Morris, S. L., Weymann, R. J., Savage, B. D., \& Gilliland,
R. L. 1991, \apj, 377, L21

\reference Miralda-Escud\'e, J., Cen, R. Y., Ostriker, J. P., \&
Rauch, M. 1996, \apj, in press 

\reference Press, W. H., Rybicki, G. B., \& Schneider, D. P. 1993,
\apj, 414, 64


\reference Reimers, D., \& Vogel, S. 1993, A\&A 276, L13

\reference Savage, B. et~al.~1993, \apj, 413, 116

\reference Sarajedini, V., Green, R. F., \& Jannuzi, B. T. 1996, \apj,
in press

\reference Sargent, W. L. W., Young, P. J., Boksenberg, A. \& Tytler,
D. 1980, \apjs, 41

\reference Schmidt, M. 1965, \apj, 141, 1295
\reference Schmidt, M. 1963, Nature, 197, 1040

\reference Schneider, D. P., et~al.~1993, \apjs, 87, 45

\reference Songaila, A., Hu, E. M., \& Cowie, L. L. 1995, Nature, 375, 124

\reference Spinrad, H., et~al.~1993, \aj, 106, 1

\reference Spitzer, L. 1956, \apj, 124, 20

\reference Stengler-Larrea, E., et~al.~1995, \apj, 444, 64

\reference Storrie-Lombardi, L. J., McMahon, R. G., Irwin, M. J., \&
Hazard, C. 1994, \apj, 427, L13 

\reference Stocke, J. T., Shull, J. M., Penton, S., Donahue, M., \&
Carilli, C. 1995, \apj, 451 24

\reference Weymann, R. et~al.~1996, in preparation

\reference Wolfe, A. M. 1988, in QSO Absorption Lines, eds. J. C.
Blades, D. Turnshek, \& C. A. Norman, (Cambridge University Press: New
York), 297

\reference Womble, D., Sargent, W. L. W., \& Lyons, R. S. 1996, to
appear in ``Cold Gas at High Redshift'', eds. M. Bremer, H.
Rottgering, P. van der Werf, \& C. Carrilli (Kluwer) 

\reference Zhang, Y., Annino, P. \& Norman, M. L. 1995, \apj, 453, L57


\end{references}
\end{document}